\documentclass[12pt]{article}
\newcommand{\fnd}[2]{\frac{\textstyle #1}{\textstyle #2}}
\newcommand{\xrm}[1]{{\textstyle \mbox{\rm #1}}}
\newcommand{\bm}[1]{\mbox{\boldmath $#1$}}
\newcommand{\abs}[1]{\left| #1\right|}
\newcommand{\bracket}[2]{\mbox{$\left\langle #1\left| #2\right.\right
\rangle$}}
\def\fb{\mbox{$f_{0}(980)$ }}
\def\fc{\mbox{$f_{0}(1370)$ }}
\def\Dfp{\mbox{$D_{s}^{+}\rightarrow f_{0}(980)\pi^{+}$ }}
\def\Kpp{\mbox{$K^{+}\rightarrow\pi^{+}\pi^{0}$ }}
\begin{document} \baselineskip .7cm
\title{Why the \bm{f_{0}(980)} is mostly \bm{s\bar{s}}}
\author{
Eef van Beveren\\
{\normalsize\it Departamento de F\'{\i}sica, Universidade de Coimbra}\\
{\normalsize\it P-3000 Coimbra, Portugal}\\
{\small eef@teor.fis.uc.pt}\\ [.3cm]
George Rupp\\
{\normalsize\it Centro de F\'{\i}sica das Interac\c{c}\~{o}es Fundamentais}\\
{\normalsize\it Instituto Superior T\'{e}cnico, Edif\'{\i}cio Ci\^{e}ncia}\\
{\normalsize\it P-1096 Lisboa Codex, Portugal}\\
{\small george@ajax.ist.utl.pt}\\ [.3cm]
\and
Michael D. Scadron\\
{\normalsize\it Physics Department}\\
{\normalsize\it University of Arizona}\\
{\normalsize\it Tucson, AZ, 85721 USA}\\
{\small scadron@physics.arizona.edu}\\ [.3cm]
{\small PACS number(s): 14.40.Cs, 12.39.Pn, 13.75.Lb}\\ [.3cm]
{\small hep-ph/0009265}}
\date{\today}
\maketitle

\begin{abstract}
We exploit the $W$-emission process to study the measured weak decay
\Dfp. We conclude that the scalar \fb meson is mostly $s\bar{s}$, which
is supported by different model studies.
\end{abstract}
\clearpage

Of recent interest is the question of whether the scalar meson \fb consists
mostly of nonstrange or of strange quarks \cite{DLS99,T95,TR96,IS96,BR99a}.

In the naive $q\bar{q}$ picture, one can treat the isoscalar \fb as
either mostly nonstrange and nearly degenerate with the isovector
$a_0$(980) --- like the vector pair $\omega$(782) and $\rho$(770) --- or take
the \fb as mostly $s\bar{s}$, in analogy to the close by, almost pure
$s\bar{s}$ vector $\phi$(1020). We prefer the latter interpretation, which is
also obtained in the unitarized quark/meson model of Ref. \cite{B86}
(see below). Another possibility is the four-quark scheme \cite{Jaf77,BFS00},
in which both the \fb and $a_0$(980) have substantial nonstrange and strange
components of comparable magnitude.
We may study the prior question from the recently measured weak decay rate
\cite{PDG00}

\begin{equation}
\Gamma\left(\Dfp\right)\; =\; (2.39\pm 1.06)\cdot 10^{-14}
\; GeV
\;\;\; ,
\label{GDfp}
\end{equation}

\noindent
due to the branching ratio of $(1.8\pm 0.8)$ percent and the lifetime of
$(0.496\pm 0.01)\cdot 10^{-12}$ $s$.

Given the rate of equation (\ref{GDfp}), we may express the decay amplitude
magnitude as ($p=732\; MeV$ and $m_{D_{s}^{+}}=1.9686\; GeV$)

\begin{equation}
\abs{{\cal M}\left(\Dfp\right)}\; =\;
m_{D_{s}^{+}}\sqrt{\fnd{8\pi\Gamma}{p}}\; =\;
(1.78\pm 0.40)\cdot 10^{-6}
\; GeV
\;\;\; .
\label{MDfp}
\end{equation}

Assuming that the \fb is pure $s\bar{s}$, we may account for the
experimental amplitude of equation (\ref{MDfp}) via the $W^{+}$-emission
graph of figure (\ref{Wemission}).

\begin{figure}[ht]
\Large
\begin{center}
\begin{picture}(300,150)(-150,-100)
\put(-100,-50){\line(1,0){200}}
\put(-100,-90){\line(1,0){200}}
\put(4,-5){\line(1,0){96}}
\put(4,-5){\line(4,1){96}}
\put(-102,-50){\makebox(0,0)[rc]{$c$}}
\put(102,-50){\makebox(0,0)[lc]{$s$}}
\put(-102,-90){\makebox(0,0)[rc]{$\bar{s}$}}
\put(102,-90){\makebox(0,0)[lc]{$\bar{s}$}}
\put(-20,-22){\makebox(0,0)[rc]{$W^{+}$}}
\put(102,-5){\makebox(0,0)[lc]{$\bar{d}$}}
\put(102,19){\makebox(0,0)[lc]{$u$}}
\put(-125,-70){\makebox(0,0)[rc]{$D_{s}^{+}$}}
\put(125,-70){\makebox(0,0)[lc]{$f_{0}(980)$}}
\put(125,7){\makebox(0,0)[lc]{$\pi^{+}$}}
\end{picture}
\end{center}
\normalsize
\caption[]{Contribution of $W^{+}$ emission to the process \Dfp}
\label{Wemission}
\end{figure}

\noindent
This predicts, using the values given in \cite{PDG00} for the Fermi
coupling constant, $G_{F}$, the cosine of the phase from the CKM weak
mixing matrix, $c_{1}$, and the pion decay constant, $f_{\pi}$,

\begin{equation}
\abs{{\cal M}\left(\Dfp\right)}_\xrm{\footnotesize $W^{+}$-emission}
\; =\;\fnd{G_{F}c_{1}^{2}}{\sqrt{2}}f_{\pi}
\left( m_{D_{s}}^{2}\; -\; m_{f_{0}}^{2}\right)
\;\approx\; 2.13\cdot 10^{-6}
\; GeV
\;\;\; .
\label{MDfpW}
\end{equation}

We use here for the Fermi coupling constant, $G_{F}$, the value given
in \cite{PDG00}, as well as for $c_{1}$, the cosine of the Cabibbo angle,
$\Theta_{1}$ ($\Theta_{1}\approx 12.8^{\circ}$, {\it i.e.}, $V_{ud}=0.975$).
We moreover assume $V_{cs}=V_{ud}$ at the accuracy of this paper.
For the pion formfactor, $f_{\pi}$, we take the value $93\; MeV$.

Not only is equation (\ref{MDfpW}) compatible with the data of
equation (\ref{MDfp}), but the analogous $\Delta I=3/2$ $W^{+}$-emission
process \Kpp is, too \cite{Sca84}:

\begin{equation}
\abs{{\cal M}\left(\Kpp\right)}_\xrm{\footnotesize $W^{+}$-emission}
\; =\;\fnd{G_{F}s_{1}c_{1}}{2\sqrt{2}}f_{\pi}
\left( m_{K}^{2}\; -\; m_{\pi}^{2}\right)
\;\approx\; 1.9\cdot 10^{-8}
\; GeV
\;\;\; ;
\label{MDfpWth}
\end{equation}

\noindent
the data being \cite{PDG00} $(1.834\pm 0.007)\cdot 10^{-8}\; GeV$.

By comparing formula (\ref{MDfp}) with the prediction of formula
(\ref{MDfpW}), we could infer a small scalar mixing
of $n\bar{n}$ and $s\bar{s}$, which is close to the theoretical
estimates. Since $\bracket{\sigma}{f_{0}}=0$, a mixing angle,
$\phi_{s}$, of about $14^{\circ}$ \cite{Nap98} to $20^{\circ}$
\cite{Sca82,DS98}, predicts an amplitude ratio

\begin{equation}
\fnd{\abs{{\cal M}\left(\Dfp\right)}}
{\fnd{G_{F}c_{1}^{2}}{\sqrt{2}}f_{\pi}
\left( m_{D_{s}}^{2}\; -\; m_{f_{0}}^{2}\right)
\cos\left(\phi_{s}\right)}
\;\approx\; 0.9\pm 0.2
\;\;\; ,
\label{mixing}
\end{equation}

\noindent
which certainly shows good agreement between the experimental decay amplitude
(~\ref{MDfp}) and the $W^{+}$-emission result (~\ref{MDfpW}).

As for earlier interpretations of \fb being mostly $s\bar{s}$, we refer
to the 1982 and 1986 papers of the present authors, \cite{Sca82} and
\cite{B86}. In the latter case the nonstrange and strange low-lying scalar
mesons are modelled jointly with the pseudoscalars and vectors including
mesons which contain $c$ and $b$ quarks. The resulting four-parameter model,
which fits very well low-energy $S$- and $P$-wave meson-meson scattering
cross
sections as well as the bound state and resonance positions, predicts
two complete scalar-meson nonets, one below and one above 1 {\it GeV}.

Also we note that the mostly $s\bar{s}$ structure of the \fb is stressed in
the more recent work of T\"{o}rnqvist \cite{T95} and of T\"{o}rnqvist and
Roos
\cite{TR96}, the latter of which focuses on the nonstrange $\sigma
(400-1200)$.
Let us recall here the DM2 data \cite{DM2} measuring the nonstrange large
$\sigma$ {\it bump} in the $\pi\pi$ mass distribution for
$J/\psi\rightarrow\omega\pi\pi$, but only a small \fb {\it pimple},
which shows the smallness of the nonstrange content of the \fb meson.

Also the recent Fermilab E791 collaboration \cite{E791} suggests that the
\fb is mostly $s\bar{s}$, but that the \fc is mostly nonstrange --- as hinted
by the naive quark model --- since $D_s^+\rightarrow K^+K^-\pi^+$ is not
seen.
It should be noted that the same is predicted in the unitarized meson model
of
Ref.~\cite{B86} (see also Ref.~\cite{BR99a}), contrary to e.g.\
Refs.~\cite{T95} and \cite{TR96}, which interpret the \fc as mostly
$s\bar{s}$. This discrepancy between two unitarized models is quite striking
and deserves a little more attention.

There are clearly too many scalar mesons below roughly 1.5 GeV than those
needed to constitute one scalar nonet. Therefore, models that accept the
light
($<$ 1GeV) scalars as genuine mesonic states with some kind of quark
substructure --- irrespective of the \em precise \em configuration --- and
not
just as two-meson resonance effects due to strong $t$-channel exchanges and
nearby thresholds (see e.g.\ Ref.~\cite{IS96}), normally consider the scalars
above 1 GeV as excited states. However, in the unitarized models of
Refs.~\cite{BR99a,B86} and \cite{T95,TR96} all scalar mesons below 1.5 GeV
originate in one and only \em bare \em scalar $q\bar{q}$ nonet, somewhere
in the mass region 1.3--1.5 GeV. Here, ``bare'' refers to the model situation
where the coupling to two-meson states is switched off. Now, when this
coupling is set to the model value which fits the data, the bare
scalar spectrum gets deformed and even extra states show up, as for example
the
$f_0(400-1200)$ or $\sigma$ meson, the $a_0(980)$, and, of course, the \fb
(see Ref.~\cite{BR99a} for more details).

In the model of Ref.~\cite{BR99a,B86}, the extra states constitute a complete
light nonet, including a $K_0^*(727)$ or $\kappa$ meson, while the pattern of
masses of the heavier scalars, which are nonetheless dominantly radial ground
states, remains largely unaltered.
For instance, the mostly strange $f_0$ stays around 1.5 GeV, while
the mainly nonstrange $f_0$ settles close to 1.3 GeV, apparently quite
compatible with the \fc. On the other hand, Ref.~\cite{T95,TR96} find
neither a light $\kappa$, nor the $f_0(1500)$, and interpret the \fc as
mostly $s\bar{s}$. We do not wish to enter here into a detailed discussion on
which interpretation is favored by experiment, and refer to Ref.~\cite{BR99a}
for an analysis of observed decay modes of the \fc and $f_0(1500)$ that, we
believe, support our view. In any case, this is evidently the most attractive
scenario, in which there is a complete doubling of scalar states, but keeping
the internal mass pattern of the nonets intact. To conclude the discussion of
the \fc, we would like to mention a very recent lattice calculation of scalar
quarkonium masses \cite{LW00}, which is in agreement with our interpretation
of
the \fc and $f_0(1500)$.

In conclusion and summarizing, by studying the $W$-emission process to
describe
the weak decay \Dfp,
we are able to reproduce the recently measured decay rate, provided that
the \fb is assumed to be mostly $s\bar{s}$. A mixing angle of about $14^
\circ$,
corresponding to a small $n\bar{n}$ admixture, is predicted. This
interpretation of the \fb is in agreement with previous work of the authors
\cite{DLS99,BR99a,Sca82,B86}, and also with other model studies
\cite{T95,TR96,Nap98,IIITT97,BFSS98,OOP99}. Finally, it is pointed out that
a very recent experiment \cite{E791} suggests the \fc is mostly $n\bar{n}$,
as
predicted by the unitarized meson model of Ref.~\cite{BR99a,B86}.
\vspace{0.3cm}

{\bf Acknowledgement}: We wish to thank Prof. R.L. Jaffe for useful comments on
the first draft of our manuscript.
This work is partly supported by the
{\it Funda\c{c}\~{a}o para a Ci\^{e}ncia e a Tecnologia}
of the {\it Minist\'{e}rio da
Ci\^{e}ncia e da Tecnologia} \/of Portugal,
under contract numbers
PESO/\-P/\-PRO/\-15127/\-99,
POCTI/\-35304/\-FIS/\-2000,
and
CERN/\-P/\-FIS/\-40119/\-2000.


\clearpage

\begin{thebibliography}{11}

\bibitem{DLS99}               
R.~Delbourgo, D.~Liu and M.D.~Scadron,
Phys.\ Lett.\ {\bf B446}, 332 (1999).

\bibitem{T95}                 
Nils~A.~T\"{o}rnqvist,
Zeit.\ Phys.\ {\bf C68}, 647 (1995).

\bibitem{TR96}                 
Nils~A.~T\"{o}rnqvist and Matts~Roos,
Phys.\ Rev.\ Lett.\ {\bf 76}, 1575 (1996).

\bibitem{IS96}                 
N.~Isgur and J.~Speth,
Phys.\ Rev.\ Lett.\ {\bf 27}, 2332 (1996).

\bibitem{BR99a}
E.\ van Beveren and G.\ Rupp, Eur.\ Phys.\ J.\ {\bf C10}, 468 (1999).

\bibitem{B86}                 
E.~van~Beveren, T.~A.~Rijken, K.~Metzger, C.~Dullemond, G.~Rupp, and
J.~E.~Ribeiro, Zeit.\ Phys.\ {\bf C30}, 615 (1986).

\bibitem{Jaf77}
R.\ Jaffe, Phys.\ Rev.\ {\bf D15}, 267 (1977).

\bibitem{BFS00}
D.\ Black, A.~H.~Fariborz, and J.~Schechter, arXiv:hep-ph/0008246.

\bibitem{PDG00}
Particle Data Group, D.E.~Groom \em et al, \em
Eur.\ Phys. J.\ {\bf C15}, 1 (2000).

\bibitem{Sca84}                 
M.D.~Scadron,
Phys.\ Rev.\ {\bf D29}, 1375 (1984);

here, vacuum saturation rather than $W^{+}$ emission was used.

\bibitem{Nap98}                 
M.~Napsuciale,
{\it Scalar meson masses and mixing angles in a $SU(3)\times SU(3)$
linear sigma model},
arXiv:hep-ph/9803396.

\bibitem{Sca82}                 
M.D.~Scadron,
Phys.\ Rev.\ {\bf D26}, 239 (1982).

\bibitem{DS98}                 
R.~Delbourgo and M.D.~Scadron,
Int.\ J.\ Mod.\ Phys.\ {\bf A13}, 657 (1998).

\bibitem{DM2}                 
DM2 collaboration, J.~Augustin et al,
Nucl.\ Phys.\ {\bf B320}, 1 (1989).

\bibitem{E791}                 
Fermilab E791 collaboration, E.M.~Aitala et al,
{\it Study of the $D_{s}^{+}\rightarrow\pi^{-}\pi^{+}\pi^{+}$ decay
and measurement of $f_{0}$ masses and widths},
arXiv:hep-ex/0007027.

\bibitem{LW00}
W.\ Lee and D.\ Weingarten, Phys.\ Rev.\ {\bf D61}, 014015 (2000).

\bibitem{IIITT97}
S.~Ishida, M.~Ishida, T.~Ishida, K.~Takamatsu, and T.~Tsuru, Prog.\ Theor.\
Phys.\ {\bf98}, 621 (1997).

\bibitem{BFSS98}
D.\ Black, A.~H.~Fariborz, F.~Sannino, and J.~Schechter, Phys.\ Rev.\
{\bf D58}, 054012 (1998).

\bibitem{OOP99}
J.~A.~Oller, E.~Oset, and J.~R.~Pel\'{a}ez, Phys.\ Rev.\ {\bf D59}, 074001
(1999).

\end{thebibliography}
\end{document}